\documentclass[journal]{vgtc}                     

\onlineid{1548}

\vgtccategory{Research}

\vgtcpapertype{evaluation}

\title{\textit{``It looks sexy but it's wrong.''} Tensions in creativity and accuracy using genAI for biomedical visualization}

\author{%
  \authororcid{Roxanne Ziman}{},
  \authororcid{Shehryar Saharan}{},
  \authororcid{Ga{\"e}l McGill}{},
  \authororcid{Laura Garrison}{0000-0001-7134-2006}
}

\authorfooter{
  \item
  	Roxanne Ziman and Laura Garrison are with Univ. of Bergen.\\
    E-mail: \{roxanne.ziman\,$|$\,laura.garrison\}@uib.no\,.
  \item
  	Shehryar Saharan is with Univ. of Toronto.
  	E-mail: s.saharan@utoronto.ca.
  \item
  	Ga{\"e}l McGill is with Harvard Medical School and Digizyme.\\
  	E-mail: mcgill@digizyme.com.
}

\abstract{
 We contribute an in-depth analysis of the workflows and tensions arising from generative AI (genAI) use in biomedical visualization (BioMedVis). Although genAI affords facile production of aesthetic visuals for biological and medical content, the architecture of these tools fundamentally limits the accuracy and trustworthiness of the depicted information, from imaginary (or fanciful) molecules to alien anatomy. Through 17 interviews with a diverse group of practitioners and researchers, we qualitatively analyze the concerns and values driving genAI (dis)use for the visual representation of spatially-oriented biomedical data. 
 We find that BioMedVis experts, both in roles as developers and designers, use genAI tools at different stages of their daily workflows and hold attitudes ranging from \textit{enthusiastic adopters} to \textit{skeptical avoiders} of genAI. 
 In contrasting the current use and perspectives on genAI observed in our study with \textit{predictions} towards genAI in the visualization pipeline from prior work, we refocus the discussion of genAI's effects on projects in visualization in the here and now with its respective opportunities and pitfalls for future visualization research. 
 At a time when public trust in science is in jeopardy, we are reminded to \textit{first do no harm}, not just in biomedical visualization but in science communication more broadly. Our observations reaffirm the necessity of human intervention for empathetic design and assessment of accurate scientific visuals. Supplemental study materials are available at \href{https://osf.io/mbw86/?view_only=e087ab5b90a6474abec7bfc42cd2b105}{\texttt{https://osf.io/genaixbiomedvis/}}.
}


\teaser{
  \centering
  \includegraphics[width=\textwidth]{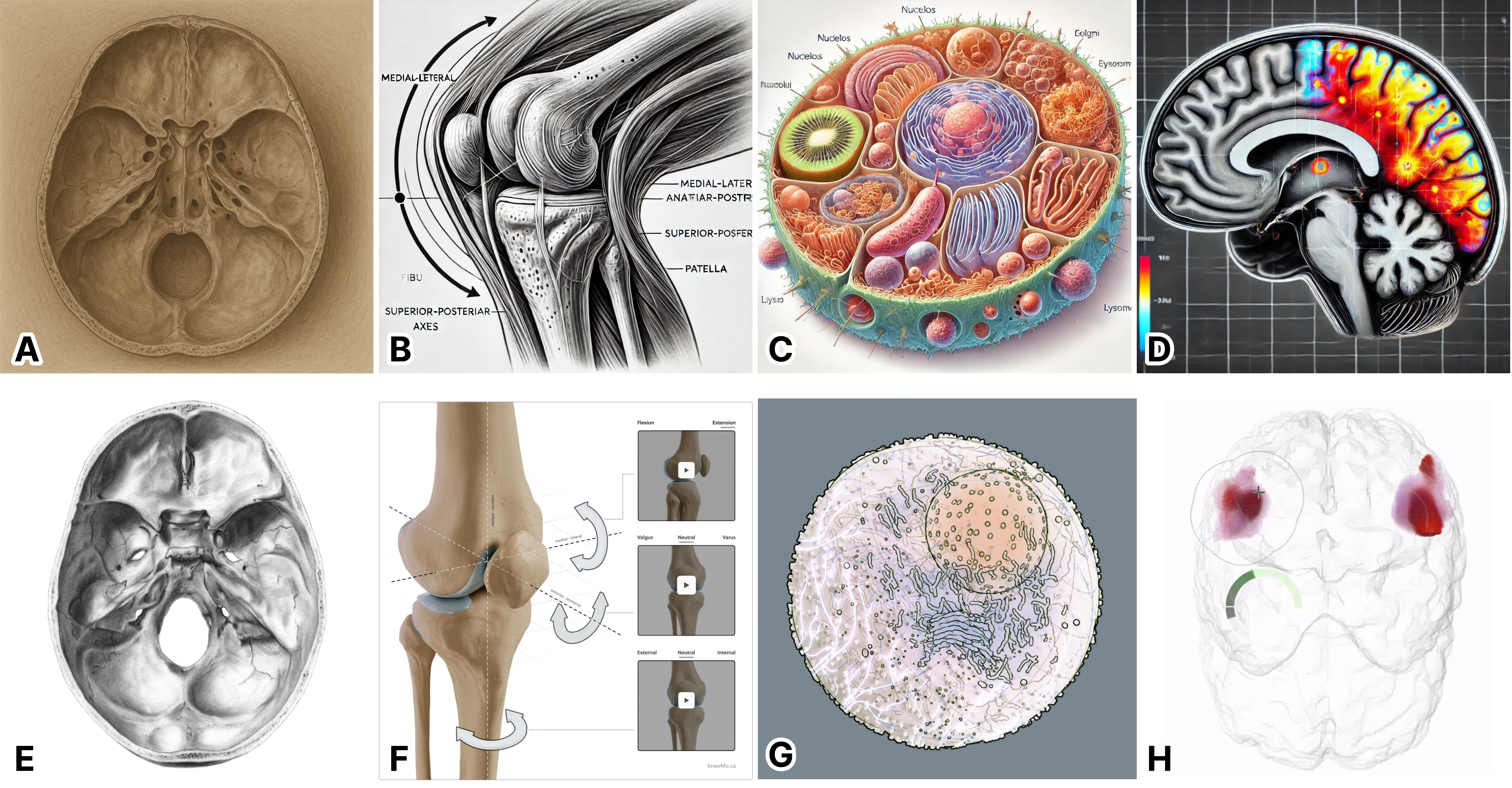}
  \caption{\textbf{A--D}: Visual outputs from GPT‑4o~(A) or DALL-E 3 (B--D), the predecessor to the most used text-to-image genAI model amongst interviewees, with prompts for biomedical topics (generated mid-March 2025). While employing conventional biomedical visualization techniques with highly aesthetic rendering, these visuals remain factually wrong or incapable of realizing the desired image to varying degrees (see supplementary material for details). \textbf{E--H}: Corresponding intended outputs produced by biomedical visualization experts. }
  \label{fig:teaser}
}

\graphicspath{{figs/}{figures/}{pictures/}{images/}{./}} 

\usepackage{tabu}                      
\usepackage{booktabs}                  
\usepackage{lipsum}                    
\usepackage{mwe}                       
\usepackage{enumitem}

\usepackage{mathptmx}                  

\usepackage{soul}
\usepackage{soulpos}
\usepackage[dvipsnames]{xcolor}



\definecolor{Dev}{HTML}{A474E4}
\definecolor{Comm}{HTML}{E2B000}

\definecolor{DevHL}{HTML}{DAC2FA}
\definecolor{CommHL}{HTML}{FEE796}
\definecolor{CutHL}{HTML}{87CEEB}

\newcommand{\dev}[1]{{\setlength{\fboxsep}{0pt}\colorbox{DevHL}{#1}}} 
\newcommand{\comm}[1]{{\setlength{\fboxsep}{0pt}\colorbox{CommHL}{#1}}}

\newcommand{\revision}[1]{{\color{black}{#1}}}

\begin{document}



\firstsection{Introduction}
\maketitle
%
A sepia-toned hand-shaded skull reminiscent of da Vinci, albeit with misplaced or missing holes. 
A beautifully-rendered illustration of one axis of knee biomechanics with confusing, misspelled labels and incorrect tendon insertions. A kiwi fruit embedded in a ``didactic'' colourized human cell. A human brain in simulated functional magnetic resonance imaging data with alien folds and psychedelic activation patterns. Shown in \cref{fig:teaser}A--D, these are but a few scenarios encountered in our exploration of prompt-based generative AI (genAI) tools that include ChatGPT and DALL-E~\cite{openai_chatgpt_2024} in biomedical visualization (BioMedVis). A trained eye may recognize the (sometimes subtle) inaccuracies in these otherwise beautiful visuals. Some issues are obvious: the infamously well-endowed rat in a now-retracted article published in \textit{Frontiers in Cell Development Biology}~\cite{knapton_ai-generated_2024} is difficult to forget. 
\revision{In light of} GPT-4o Image Generation's public release at the time of this writing~\cite{openai2025gpt4o}, visuals produced by genAI often look polished and professional enough to be mistaken for reliable sources of information. 
This illusion of accuracy can lead people to make important decisions based on fundamentally flawed representations, from a patient without such knowledge or training inundated with seemingly accurate AI-generated ``slop'', to an experienced clinician who makes consequential decisions abut human life based on visuals or code generated by a model that cannot guarantee 100\% accuracy. Troubling on a larger scale is the possibility that the flood of inaccurate, yet compelling, AI-generated visuals may erode trust in images that are carefully and accurately produced by human experts. 

A flurry of recent research has explored tooling to integrate production-ready natural language and diffusion models in visualization workflows~\cite{mittenentzwei_ai-assisted_2024, ye2024generative, buzzaccarini_promise_2024, kim_generative_2024, basole_generative_2024}, and investigated content creator perspectives in fields outside such as fine art~\cite{bird_artists_2024, li_user_2024} or creative knowledge work more broadly~\cite{woodruff_how_2024}. Yet comparatively little work has engaged directly within visualization to understand individuals' current attitudes and use of genAI in their workflows. Schetinger et al.~\cite{schetingerdoom2023} speculate on the uses and risks of genAI tools in the general visualization pipeline, but their work does not reflect \textit{current} use in light of the seismic changes in model accuracy, nor does their study focus on the unique considerations that come into play in a field like BioMedVis. 

\revision{BioMedVis is a specialty of scientific visualization focused on  strategies that facilitate exploratory analysis and communication of complex phenomena in biology and medicine~\cite{preim_visualization_2007,furmanova2024biomedvis,kaufman_overview_2014}.} Encompassing competencies ranging from computer science to design and fine arts, BioMedVis is of particular interest for evaluating genAI for its focus in depicting complex spatial relationships over broad scales of both space and time. Unlike statistical graphics, even the simplest biomedical visualization comprises myriad interconnected variables, \revision{from subcellular to body-level anatomy (e.g., showing molecular malformations that underlie cancer to the movement and growth of tumours in organs of the body~\cite{garrison_exploration_2021})} where relationships must be \textit{spatially maintained} \revision{while adapting visual abstraction for different audiences~\cite{christiansen_building_2022, garrison_exploration_2021}}. In representing such data, BioMedVis provides a case study for the careful traversal of the tight relationship between science and figurative visuals. This is a relationship which genAI, with its ease of producing slick, believable visuals, is uniquely placed to impact. 

To understand the current uses and attitudes around genAI in this field, we interviewed 17 \textit{creative knowledge workers}~\cite{woodruff_how_2024} in BioMedVis. 
We report on points in the visualization pipeline where the BioMedVis expert is already using and affected by genAI in their daily work. We describe participants' considerations when designing visualizations for biology and medicine that require: empathetic and experiential design, hyper-specialization to render spatial relationships at a high fidelity, and creative execution. However, beneath these broad themes that seem to point towards disuse of genAI tools, we see a multifaceted decision landscape for genAI that includes career stage, discipline, role, and closely-held personal values regarding the nature of technology and craft. Navigating this landscape provides insights into the factors guiding those who \textit{enthusiastically adopt} versus those who are \textit{skeptical and avoidant of} genAI, and alerts the broader visualization community to opportunities in supporting BioMedVis experts, particularly those embedded outside the computer science context, in the daily work that is, for better or worse, likely to be infused with genAI.

\revision{\textbf{Our primary contribution is an in-depth qualitative analysis on the use and attitudes towards genAI in BioMedVis research and practice}, offering an exploration through our interviews of the social, labour, and economic factors that drive individuals towards genAI (dis)use. We investigate participants' perspectives towards genAI alongside broader reflections on its transformative role 
in science communication and visualization, through its continuously evolving technology and social pressures.} Guided by Walford's \textit{Data Aesthetics}~\cite{walford_data_2020}, we critically reflect on human use of genAI to \textit{craft rather than reveal truth} in communicating visual scientific stories. 

\textbf{Our secondary contribution updates current knowledge for the use of genAI in the visualization pipeline} relative to the speculations posed by Schetinger et al.~\cite{schetingerdoom2023} shortly after the public release of ChatGPT in late 2022, and introduces further considerations for use with complex biomedical data. 
As genAI technology continues its rapid evolution~\cite{zhang_mm-llms_2024}, we provide a timely and focused discussion on genAI in researching and producing visual, spatially-oriented scientific imagery.

\section{BioMedVis Context}
\label{sec:background}

BioMedVis is essential to facilitate domain expert exploration and analysis of biomedical structures and events, e.g., skeletal anatomy, knee biomechanics, cellular activity, or neuronal activation \revision{(see~\cref{fig:teaser}E--H)}. In communication, biomedical visualizations are integral to 
atlases used for clinical training~\cite{daston1992image}, appear as figures and cover art to showcase science in academic journals, are employed in multimedia learning aids in education and research\revision{~\cite{jenkinson2018molecular}}, and serve as informative illustrations or animations that bridge medical science to patients and the broader \revision{public~\cite{garrison_exploration_2021}}. \revision{In the following, we introduce the data sources, necessary skill set, and key pipeline stages for BioMedVis.}


\paragraph{Data}
BioMedVis often aims to preserve and emphasize spatial relationships in data that vary broadly over space and time~\cite{preim_visualization_2007,furmanova2024biomedvis,kaufman_overview_2014}. 
Nano- and microscale data include molecular structure, e.g., cryo-electron tomography data, and dynamics data as well as omics, e.g., genomics data. Mesoscale data describing cell- and tissue-level structure and behaviours include, e.g., histological image sections or live microscopy methods to flag structures of interest. Organ- and body-scale data sources include computed tomography (CT) and positron emission tomography (PET). For further details on BioMedVis data sources we refer the reader to Garrison et al.'s survey~\cite{garrison_trends_2022}. Generally sourced from human patients, these data come with stringent requirements for secure and ethical use~\cite{hsiao_establishing_nodate}.

\paragraph{Skill set}
The BioMedVis skill set is diverse and spans multiple disciplines beyond visualization to include: computer science, data science, UI/UX design, design, fine art, communication, and marketing~\cite{preim_visualization_2007, kaufman_overview_2014, christiansen_building_2022}. Many professionals have expertise in the biological and/or medical domain, typically through formal study~\cite{association_of_medical_illustrators_learn_nodate, garrison_exploration_2021}. 


\paragraph{Pipeline}
\revision{The BioMedVis pipeline we describe synthesizes prior literature on the general visualization pipeline~\cite{dos_santos_gaining_2004} and the scientific visualization framework~\cite{zhang_framework_2023}. This synthesis is also reflective of our own experiences as BioMedVis researchers and practitioners, and those of our study participants, to follow three overarching stages:}


\begin{enumerate}[nosep]
    \item \textbf{Research and ideation}: Identify opportunities for, or orient to, a new project. Ideate or draft potential concepts~\cite{zhang_framework_2023}, or clean and process data in preparation for visualization~\cite{dos_santos_gaining_2004, basole_generative_2024}.
    \item \textbf{Implementation/production}: Design or program a novel visualization through mapping and rendering geometric/scalar data, often synthesizing multiple such sources, into the final image(s)~\cite{zhang_framework_2023, garrison_exploration_2021, dos_santos_gaining_2004, basole_generative_2024}. Biological and medical phenomena are often characterized by spatial data~\cite{preim_visualization_2007,odonoghue_visualization_nodate,kaufman_overview_2014}. These spatial representations may be figurative, i.e., a close replication of the source material, and vary in style and degree of interactivity~\cite{zhang_framework_2023}. Concept-driven figurative visuals, e.g.,~\cref{fig:teaser}E and \cref{fig:teaser}F, are usually produced by designers or illustrators~\cite{goodsell_filling_2007, christiansen_building_2022} with the appropriate visual abstraction for the intended story and audience, or when data are absent or lack the necessary fidelity. Data-driven figurative representations, as in \cref{fig:teaser}H, are typically produced by visualization developers~\cite{mindek_visualization_2018, garrison_exploration_2021}, although there is a natural midpoint where skill set and visual style meet, as in \cref{fig:teaser}G. Termed \textit{illustrative visualization}~\cite{rautek_illustrative_2008, sousa_illustrative_nodate}, this discipline draws inspiration from traditional illustration techniques to render scientific data. On the opposite end of the representational spectrum, visuals may be fully abstracted from their spatial context~\cite{leitte_visualization_2014, christiansen_building_2022}. Regardless of the representational approach, a developer or designer may collect and combine multiple, heterogeneous data sources to build the final visualization~\cite{jenkinson2018molecular,garrison_exploration_2021}. 
    \item \textbf{Dissemination}: Refinement, final delivery, and dissemination, e.g., publishing, of the work~\cite{zhang_framework_2023, garrison_exploration_2021}. Across all stages are numerous touch-points for review and discussion~\cite{zhang_framework_2023, garrison_exploration_2021, dos_santos_gaining_2004, basole_generative_2024}. 
\end{enumerate}

BioMedVis, then, is a professional and research practice that bridges science and art through complex, multidimensional forms. In this work, we explore how genAI has modulated this tension from both a broad perspective and a detailed investigation into BioMedVis daily workflows with genAI. 

\section{Related Work}
Recent years have seen various scientific and creative fields grappling with the impact and future of genAI. To the best of our knowledge, our work is the first detailed investigation of genAI use in BioMedVis.

\paragraph{GenAI in Creative and Scientific Pursuits}
Several qualitative studies have sought to understand perceptions of genAI in creative 
domains. Vimpari et al.~\cite{vimpari_adapt-or-type_2023} summarize a practical ``adapt-or-die'' perspective held by designers in the gaming industry, highlighting copyright issues, potential job loss, and shifting roles and skill sets as very likely. This echoes fears and sentiments amongst professionals across creative and communications industries~\cite{wolters_animation_2024}. In user experience design, professionals imagine the potential for human-AI collaboration while elevating the unique human-human relational aspects that are a cornerstone of this domain~\cite{li_user_2024}. 
Palani et al.~\cite{palani_evolving_2024} similarly explore human-AI collaboration by ``orchestrating with genAI'' throughout the creative process, highlighting tensions and opportunities in genAI use.

Bird~\cite{bird_artists_2024} explores challenges and taboos experienced by creatives working with genAI, themes which resonate in our work. Woodruff et al.~\cite{woodruff_how_2024} organized workshops to understand attitudes toward genAI amongst knowledge workers, those who perform specialized, knowledge-based work~\cite{galbraith_kenneth_new_2007}, across multiple professional areas adjacent to BioMedVis, such as education, journalism, and mental health. Primary concerns raised include the risk of increased disinformation~\cite{menz_health_2024} along with the proliferation of low-quality content, and dehumanization, which removes the human-centred quality of this work. These themes echo in BioMedVis through the development and dissemination of materials critical to scientific analysis and communication. 

The academic and scientific research community at large is similarly grappling with the challenge of how to preserve scientific integrity with the advent of genAI tools in research~\cite{alasadi2023generative, blau2024protecting}. Summarizing a body of work, writing manuscripts with correct grammar, and citing relevant work for a topic are all proposed uses for genAI in different scientific communities, even within visualization. In contrast, our interview study focuses less on these text-to-text aspects in the modern scientific process and instead on the means of visualization design and development in a research or production context for topics in biology and medicine. Our emphasis on the need to preserve spatial relations in visuals more figurative than abstract introduces a different set of constraints than previously identified for genAI-assisted workflows.

\paragraph{GenAI in Visualization}
Visualization-focused studies have explored both attitudes and feasibility of genAI in data visualization. Recent surveys by Basole and Major~\cite{basole_generative_2024} and Ye et al.~\cite{ye2024generative} outline the creative and automated tasks that can be enhanced by implementing genAI tools across the general visualization pipeline, along with related challenges. For instance, the authors identify automation tasks as valuable instantiations of genAI adoption, a sentiment we heard echoed in our own study. Schetinger et al.~\cite{schetingerdoom2023} explore attitudes, emotions, opportunities, and threats of genAI adoption throughout the general information visualization pipeline. Their work strongly inspired our study design. Yet our work follows over a year after this study---a consequential period of time given the rate of improvement of each new LLM or diffusion model public release~\cite{zhang_mm-llms_2024}. Finally, these various studies explore in a limited way the concerns of spatial data visualization that is often inherent to projects in biomedical visualization. In contrast, we centre our questions around the rendering of spatial phenomena.  

Within biomedical visualization, Kim et al.~\cite{kim_generative_2024} experiment with OpenAI's DALL-E 3~\cite{openai_chatgpt_2024} to generate images of biomedical subject matter, including cell cultures, histological slides, and medical diagnostic imaging, e.g., X-rays. The authors laud the genAI tool's ability to render ``convincing'' and 
``visually compelling'' scientific visualizations, but note the gross inaccuracies, such as misrepresenting anatomy or including irrelevant subject matter, that limits its use in scientific visualization. Two other studies focus on medical illustrations, one related to aesthetic surgical applications~\cite{buzzaccarini_promise_2024}, the other on corneal surgical procedures~\cite{moin_assessment_2024}. Both produce inaccurate results. Of note, the authors of these three studies appear to be primarily situated within medical and clinical contexts, and conclude that collaboration between medical experts and AI developers will be necessary for developing these tools. \textbf{We emphasize the crucial and (currently) absent voice of the BioMedVis expert, whose experience and training in human-centred storytelling is an essential bridge between aesthetics and accuracy, which is missing in these projects.} Through this work, we profile the values and skills that BioMedVis experts bring to these conversations and surface potential opportunities for collaboration. 
We complement other voices from within the community that have explored genAI tools to support the narrative of a patient-centred medical visualization. For example, Mittenentzwei et al.~\cite{mittenentzwei_ai-assisted_2024} develop and evaluate a genAI-assisted workflow to produce photo-realistic characters representing patients in data-driven medical stories. This workflow is time-efficient and democratizes the otherwise specialized process of character modelling, rigging, and animation. Moreover, their evaluation showed a receptiveness to AI-generated images amongst viewers.
In our study, we explore a broader range of illustrative biomedical visualization techniques used by BioMedVis experts that capture other potential workflows and uses of genAI.

\section{Methods}

We began this work out of a desire to understand how BioMedVis experts \textit{actually} use genAI tools, which we see as a necessary step toward developing guidelines and resources to help unite our community in the ethical and practical use of this technology. 
As BioMedVis practitioners and visualization researchers ourselves, we have observed a sense of insulation from the existential concerns around genAI since OpenAI's public release of ChatGPT in late 2022~\cite{openai_introducing_2022}. 
While some BioMedVis experts spoke of genAI as a paradigm-shifting tool, we wanted to learn \textit{if} and \textit{how} colleagues use, and are affected by, these tools in their daily lives. We were interested to learn what, if any, of the predictions described in Schetinger et al.~\cite{schetingerdoom2023} had come to pass, particularly within the BioMedVis subcommunity, which works at the margins of spatial and figurative art in science. Through semi-structured interviews with BioMedVis professionals, we elicited personal perspectives and actual ways of working, with the goal of understanding how the field is responding to genAI and its impact on the daily production of visualizations for high-stakes health contexts. For study instruments and further details, see supplementary material (\href{https://osf.io/mbw86/?view_only=e087ab5b90a6474abec7bfc42cd2b105}{\small\texttt{https://osf.io/genaixbiomedvis}}).

\subsection{Participant recruitment}
We used purposive and convenience sampling~\cite{taherdoost_sampling_2016}, which allowed us to identify 
prospective participants according to expertise and role(s) for a cross-sectional representation of our target community. We acknowledge convenience sampling because the majority of the Euro-North American community, where we focused our study, is connected through first- or second-degree relations with our author team. 
We aimed to capture a range of career stages (early, mid, and late or close to retirement); organization size (\revision{freelance or} small studio to large academic institution); and seniority (junior artists to creative directors in industry; PhD/postdocs to principal investigators in research). 
We furthermore recruited participants from different disciplines within the community; we targeted recruitment efforts to medical illustrators and designers within the Association of Medical Illustrators~\cite{association_of_medical_illustrators_learn_nodate} (from here on referred to as \comm{designers}). We also recruited from the the Workshop for Visual Computing for Biology and Medicine (VCBM)~\cite{vcbmconf} for participants with computer science or more technical backgrounds (hereon referred to as \dev{developers}). Participant profiles are illustrated in \cref{fig:profiles}.

\subsection{Interviews}
We obtained ethics approval through the University of Toronto Research Ethics Board prior to data collection. To guide our interviews we developed a semi-structured interview template divided into four sections (see supplementary material for details). This structure afforded room for both organic discussion and deeper probing into topics of interest:

\begin{enumerate}[nosep]
    \item \textbf{Introduce the participant} and their role, experience, and a brief description of their current projects and everyday tasks. 
    \item Open the discussion about genAI to elicit participants' \textbf{basic understanding of, emotions about, and attitudes toward this technology} as a broad value-add or threat to the community and practice of BioMedVis (e.g., \textit{What is genAI, in your own words?} and \textit{What kind of impact do you think genAI will have on your professional work?}). 
    \item Focus on the nuts and bolts of the BioMedVis pipeline through a \textbf{workflow activity}, designed to elicit points of discussion regarding genAI in the pipeline, following a structure similar to related interview studies in visualization~\cite{schetingerdoom2023} and UX design~\cite{li_user_2024}. Using a collaborative Miro board~\cite{miro_miro_nodate}, we walk participants through the pipeline and ask them to describe their current, potential, and resistance to genAI use, and why (e.g., \textit{Is there something that stops you from using genAI tools at x stage?}). 
    \item Explore \textbf{ethical concerns} (e.g., \textit{What are the ethical issues surrounding genAI use in BioMedVis?}). Close with participants' thoughts and predictions about its future impact on the field. 
\end{enumerate}

We conducted remote interviews via Zoom~\cite{zoom_zoom_nodate} from January--August 2024. Participants received no financial compensation for their time. We recorded video and audio with participant consent, with interviews lasting 90 minutes on average. While one researcher led the interview, the other researcher(s) navigated and memoed throughout. This enabled us to capture in-the-moment reactions and insights, about which we debriefed as a team
immediately following each interview~\cite{akbaba2023two}. We uploaded recordings to Dovetail~\cite{dovetail_dovetail_nodate} to generate written transcripts that we validated for accuracy through our analysis.

\subsection{Participants}
We interviewed 17 (11M, 6F) BioMedVis researchers, educators, and practitioners, nine of whom were based in North America and eight in Europe. Eight participants are \dev{developers}, i.e. those situated predominately in the computer science domain, while nine are \comm{designers} from the illustration/visual design domain. \revision{Three participants assume a `hybrid' role based on their work contexts, but we refer to them by their self-identified primary role.} While these categories have fuzzy boundaries, we see value in these delineations as they make explicit the professional context of each participant that impacts their exposure to, and understanding of, the technical underpinnings of genAI models. Professional expertise ranges from 3--40 and 3--25 years of experience for designers and developers, respectively. \revision{All participants primarily work digitally---designers favour creative tools such as the Adobe Suite, as well as 3D-modelling,  animation, and medical imaging software (details in suppl. materials).}
The range of domain expertise amongst participants includes molecular visualization, gross anatomy, and public health. They work with various stakeholders, including clinicians, patients, biotech companies, creatives in adjacent fields, developers, and research scientists in biomedicine. Several teach at the undergraduate and graduate levels, and have perspectives on the future of the field through the eyes of the next generation of BioMedVis experts. 

We refer to participants with \textit{pseudonyms} throughout this report to preserve anonymity and limit use of information that could lead to identifying individuals. 
\cref{fig:profiles} shows participants' pseudonyms, role (designer/developer), gender, professional setting (academia/industry) and approximate experience level. Participants are arranged according to their attitudes towards genAI, i.e., from \textit{enthusiastic adopters} to \textit{skeptical avoiders}, 
based on their responses to attitude and value-based questions in their interviews. We discuss the resulting arrangement in detail in Sec.~\ref{sec:adoptionspace}. 


\subsection{Interview Analysis}

 We analyzed data from the workflow activity held during the interview, in addition to transcripts and recordings, aiming to identify overarching themes related to genAI in BioMedVis. 
 For the workflow analysis, the first and second authors synthesized the activity held during the interviews into \textbf{temptations} and \textbf{turbulences}, that is, positive and negative aspects about genAI use, after the approaches outlined by Schetinger et al.~\cite{schetingerdoom2023} and Nadal et al.~\cite{nadal2022tac}. These uses were grounded within the three stages of the BioMedVis pipeline described in Sec.~\ref{sec:background}. 
 
 We conducted reflexive thematic analysis of the interview transcripts following Braun \& Clarke~\cite{braun_thematic_2021} to expose broader themes and patterns around participants' genAI use and attitudes. In the open coding stage, the first author applied iterative, open coding and memoing techniques to the interview transcripts until reaching a point of saturation. The first author met regularly with the research team to discuss the outcomes from open coding iterations\revision{---consolidating an initial 146 codes down to a final 138 after merging codes with similar phrasing or meaning---}or in cases where she felt there may be bias in the coding process. 
 The first author then used thematic mapping to cluster thematically-related codes around central organizing concepts and develop candidate themes based on recurrence or similarities across transcripts. For example, the codes \textit{bias in output, do no harm, sensitivity of health medical data,} and \textit{public trust} were grouped around the central organizing concept \textit{Ethics.} 
 All authors met frequently throughout this process, focusing on patterns as well as divergent views between participants, in alignment with best practices for reflexive thematic analysis that focus on qualitative rather than quantitative outcome metrics~\cite{braun_thematic_2021}. 
 We captured our own reflections from the interviews, our respective domain knowledge, experience, and perspectives, to interpret our findings and come to a consensus on the overarching themes.

\section{Findings | GenAI in the BioMedVis Pipeline}
\label{sec:results}

Through our workflow analysis we see that genAI is already, to varying degrees, altering participants' daily work. For each stage of the BioMedVis pipeline, we identify points of current genAI use. In addition, we note \textbf{temptations}, i.e., tasks or aspects of the pipeline that participants stated genAI tools help to improve, and \textbf{turbulent points}, i.e., concerns or pitfalls of genAI tool use in the BioMedVis pipeline.



\subsection{Stage 1: Research and Ideation}
The research and ideation stage initiates all BioMedVis projects: identifying the problem that needs to be solved, uncovering the user's or client needs, learning about the topic through a literature review (often about novel research), and ideation, i.e., coming up with possible solutions to the problem.
For visual development, six designers use text-to-image prompting for inspiration. Abstract and other-worldly images help designers explore different visual styles to achieve a desired aesthetic (e.g., colour palettes) or atmosphere. These outputs, \comm{Jeff} and \comm{Neil} say, can help ``jump-start'' conversations with clients or other stakeholders early in the development pipeline, and are integrated with the BioMedVis expert's trained scientific knowledge and visual storytelling techniques to create the final visuals. This aesthetic does not suit every BioMedVis workflow; \dev{Jules} described genAI images as being \textit{``like something out of a dream or a movie''} and that he \textit{``can't think of an instance to use these tools for any sort of inspirational content.''} Eight designers agree that the derivative \textit{``samey-ness of the images gets boring,''} as \comm{Frank} put it, is an issue; their work relies on clear messaging with a unique and recognizable visual style.

Text-to-text models are seen as helpful for six participants when conducting initial research on a (not necessarily novel) topic before validating the information themselves by going directly to relevant sources. \revision{Requiring a large volume of training data to perform well,} participants felt ChatGPT typically provides reasonably reliable text information about well-established scientific content (e.g., what can be found in university-level science courses~\cite{schulze2024empirical}), but is less suitable for summaries on emerging scientific research topics~\cite{schulze2024empirical}. Irrelevant or hallucinated references remain a problem, as do invented new terms, such as the \textit{``green glowing protein.''} \comm{Margaret}, who came across this term in an experiment asking ChatGPT for a summary of the green fluorescent protein, is particularly skeptical of the utility of genAI tools in her work. \dev{Ray}, \dev{Alan}, and others describe the well-known issue of \revision{\textit{``the confident idiot,''} i.e., 
a mismatch of self-awareness against real expertise}, as particularly problematic for the biomedical field, which relies on accurate information for informed decision-making in health and clinical contexts. 

\subsection{Stage 2: Implementation / Production}
This stage involves synthesizing multiple data sources, iteratively developing and refining custom algorithms or visualization techniques, and other tasks in the technical execution of the proposed solution.

For five designers and one developer, text-to-image genAI use centres around non-anatomical, non-technical visual assets, such as patient avatars and background elements. Transforming a low-resolution image or animation to high-resolution with Topaz~\cite{topazlabs}, an AI-powered image- and video-enhancing tool, also fits into this category. 

Three developers note a potential for text-to-text models to assist with synthetic data generation and some analysis, corroborating findings from related studies~\cite{pantanowitz2024synthetic, ibrahim2024generative}. \dev{Kim} sees the potential to improve clinical training by generating additional plausible variances of anatomic structures, i.e., fitting within the average range of anatomic variation, and could be observed in reality. Achieving this remains a challenge, however, while the availability of medical training data remains limited due to concerns regarding patient confidentiality and security. \dev{Alan} notes that a feasible solution could be to develop and train a model on a local or sequestered environment, although the computing power required for this would likely not be able to compete with, e.g. ChatGPT-4. However, an \textbf{adequate amount of training data for niche spatial biological and medical data is currently lacking}, and presents a significant barrier to use in any environment, sequestered or otherwise, as \dev{Ray}, \dev{Alan}, and \dev{Lois} observe. 
Two participants also expressed concerns over unknowingly generating the likeness of a real or living person in their work, or exposing a patient's data. Either scenario is seen as an enormous breach of confidentiality and ethical violation in the field, issues which participants feel are less fraught in other domains.

Ten participants identify candidate tasks for genAI in automation, including generating boilerplate or segments of code, cleaning data, and debugging. This aligns with findings in related studies~\cite{schetingerdoom2023}, and is the value proposition of tools like GitHub Copilot~\cite{github_copilot}. Depending on the participant's pipeline and preferences, this may enable automating the \textit{``boring''} stuff, i.e., tedious or not enjoyable, low-level, or human resource-intensive tasks, and is true for both designers and developers. \dev{Alan} sees roughly a ``90\%-10\% split'' in his workflow with such automation, allowing him to focus his energy on more complex or creative problems, a sentiment echoed by most other participants. \comm{Neil} sees the opportunity for genAI to speed up production altogether by: \textit{``reduc[ing] the amount of work that goes into making high-quality imagery, to reduce friction in the production process, and the goal is to eventually use AI to reduce the actual labour hours.''} 

However, skeptics like \dev{Lois} and \comm{Margaret} resist genAI tools, having already invested time to hone their process and skills. \textit{``I'm very good at coding in Python myself,''} says \dev{Lois}, \textit{``so I don't want to use ChatGPT for that.''} The joy of such work was echoed by \dev{Octavia} and \dev{Ray}. From a design perspective, \comm{Frank} comments, \textit{``I probably reach a flow state more often in my work than anything else [...] when I've already figured out all the problems in this drawing and now I'm just painting, and I'm happy, I'm free. [That] could be the kind of mindless work that a robot could be doing. But then I'd probably be sad, because I like doing that, and my work would be less joyful in some ways.''}

Participants in supervisory roles note resistance to genAI amongst some of their team, as \comm{Jeff} explains: \textit{``The [designers] certainly do [feel threatened by it], and some people [...] have been doing what they do for a very long time, they're very fast and efficient and they've got the \textbf{sense of ownership and control} -- they're not as familiar or comfortable with the genAI tools from a process standpoint.''} Participants further note a frustrating trade-off between initial time-saving convenience and time-consuming validation and correction of genAI outputs.

All participants emphasize that AI-generated content (namely, images) becomes even more ethically problematic to use later in the production stage (i.e., toward the final output) due to copyright infringement risk. While they express grave concerns about intellectual property violations that are, for the moment, baked into public genAI tools, 15 participants have a higher tolerance for personal over commercial uses of genAI, and prefer text- over image-based uses in their professional work.

\subsection{Stage 3: Dissemination}
Delivery and/or dissemination of the project occurs at this stage. Visual artifacts may support clinical training or research, patient education, or public health initiatives. 
In this stage, participants find genAI tools most useful for short, text-based elements like figure captions and metadata or translating complex technical or medical jargon to an appropriate level depending on the communication goals. As \dev{Octavia} expresses, \textit{``Can we actually use these kind of tools to translate [...] from doctors to lay people, or from visualization experts to non visualization experts? For me, this is a very exciting topic.''} \dev{Ray} adds, \textit{``One of the things I'm impressed with is how good the translation works. That's very helpful in terms of the accessibility of content [...] for the broader public.''} 

All participants agree that AI-generated elements should not be incorporated in any significant manner into final visualizations, even in editorial images that are less tied to the mandate for accuracy, and instead aim primarily to inspire. \comm{Margaret} encapsulates some of the criticism in the community towards genAI-assisted projects, noting that something designed to be engaging and interesting to audiences should still be \textit{``true to the science. It looks better when it's the actual science.''} Apart from ethics of use, participants express concerns over genAI-produced figures for scientific dissemination, noting infamous inaccurate examples of published scientific figures such as the well-endowed rat with insets of ``sterrn cells''~\cite{knapton_ai-generated_2024}.

Overall, we see participants express a prevailing sense of ``insulation'' from the impacts of genAI given its obvious limitations, while others are more open to its transformative potential in the field, despite ethical concerns. \textbf{Considering the pipeline as a whole}, 13 participants already integrate genAI to some extent in their production workflows, but primarily for tasks auxiliary to what they generally see as the core of their work. Designers avoid genAI for the manual rendering of their work, while developers resist genAI for all but the most basic code snippets when writing code for image rendering. On the whole, the designers express more favourable attitudes towards genAI tool adoption than developers, as \cref{fig:profiles} indicates. Whether developer or designer, we find participants' attitudes regarding genAI cut across a wide spectrum, from deep skepticism (\comm{Margaret}, \dev{Lois}) and distrust regarding its use in BioMedVis (\comm{Ursula}, \comm{Arthur}) to fully embracing these tools throughout the pipeline (\comm{Jeff}), as a means of supercharging creative output (\comm{Neil})--we map this landscape in \cref{fig:profiles}. Others, like \dev{Octavia} and \dev{Kim}, are enthusiastic but are not directly involved in implementation work to be using genAI tools actively. The rest fall somewhere in the middle, cautiously optimistic and keeping themselves up-to-date on genAI development in order to best understand its use and limitations. On the whole, we observe that designers appear to be slightly stronger adopters than the developers.

\section{Findings | Essentials of BioMedVis}
\label{sec:themes}

Stepping back from the BioMedVis pipeline, our reflexive thematic analysis~\cite{braun_thematic_2021} affords insights into BioMedVis experts' broader perspectives regarding genAI use. These perspectives organize into three overarching themes and associated sub-themes (\textbf{bolded} in this section) relating to accurate representation of complex spatial data, empathetic design, and balancing creative choices with accuracy and purpose.  

\subsection{Scientific accuracy forms the core of BioMedVis.}
For BioMedVis developers and designers, accuracy is a top priority that reflects both the spatial nature of the underlying data and the BioMedVis expert's commitment to ``truthful''  representation in service of science communication. GenAI in its current state is unable to achieve this benchmark. As \comm{Arthur} explains, \textit{``While it's still scraping the digital world for references it can use to generate art, it's not yet able to know the difference between the sciatic nerve and the ulnar nerve. It's just, you know, wires.''} \comm{Ursula} captures the humoured skepticism towards genAI's (in)ability to accurately produce anatomy: \textit{``Show me a pancreas, and MidJourney is like, here is your pile of alien eggs!''}

\textbf{BioMedVis often deals with niche subject matter that requires a deep understanding of 3D structures.} 
Ten participants acknowledged that custom models could be trained on only real medical data or images to produce more accurate images. However, as \dev{Anthony} states, \textit{``That's where I fail to see the utility of genAI, considering those things need to be trained on a lot of data, and lots of stuff we do is so niche.''} Moreover, quality training data is no guarantee: \comm{Margaret} describes experiments to train a custom model with a large, ``correct,'' computationally-based dataset for a common molecular structure and found the percentage of ``accurate'' AI-generated visual outputs to be low.

An interesting solution, proposed by \comm{Isaac}, is photogrammetry, a technique to capture a 360-view of a real 3D object through a series of incremental photos~\cite{linder2009digital}. This, he says, could be a way to train genAI models on real 3D anatomy, yet it would also be an intensive and laborious task considering the vast amount of anatomic subject matter and photo/image processing that would be required.

BioMedVis experts apply subject domain knowledge and 3D spatial understanding to correctly and meaningfully align visual mappings with 3D structures, as \dev{Alan}, \dev{Ray}, and \comm{Margaret} explain. \dev{Anthony} adds that the range of possible visual mappings is therefore limited as compared to other data types, and as such limits genAI's utility at this stage.

Furthermore, hyper-specialization within BioMedVis will continue to be seen as having a ``protective'' advantage, as \comm{Arthur} and \comm{Jeff} predict. The more complex, novel, or niche the subject matter, visual or technical approach, and/or modality, e.g., physical, interactive experiences (experiential design), with or without immersive tech, the less likely genAI is to serve as a collaborative partner. For example, VR-based surgical training, medical legal visualization that captures unique injuries or pathologies, and interactive exhibits, will likely remain the purview of those trained BioMedVis experts for a long time.

\subsection{Creating ``useful'' biomedical visualizations requires an empathetic, human-centred approach.}
BioMedVis experts leverage their knowledge of target audiences to create visualizations that support learning, insight-generation, and other goal-oriented tasks. Many participants stated that core to their work is close collaboration with relevant stakeholders---domain experts, clinicians, patient populations, students, or general audiences---in a project through design study~\cite{sedlmair2012design} or contracted work. 
Fluent in not only the scientific content but also a range of communication strategies, the BioMedVis expert serves as a skilled translator and visual storyteller. As \comm{Jeff} describes, \textit{``We are custom storytellers. It's listening to the client, developing that relationship [and] discussing the problem as well as the solutions. And then there's the craft of bringing that to life through visuals, and ensur[ing] [they] do the job[...].''} \dev{Jules} and \dev{Alan} echo the emphasis on relational \textit{``human connection''} as one of the irreplaceable, as well as most rewarding, aspects of the work.

\textbf{BioMedVis experts feel a strong sense of accountability for (im)proper visualization of sensitive medical and health data.}
Visualizations that support high-stakes clinical and public health decision-making leave little room for error---the consequences of which can be unforgiving.
\dev{Anthony} expresses this feeling of pressure: 
\textit{``I especially feel in the medical domain, it is just so important to know that what I'm seeing, what I'm dealing with, is correctly mapped to something I can understand. Because making wrong judgments from the data can be problematic, to say the least, right? It can be fatal in some cases.''}

Further exacerbating this issue is the (current) black-box nature of genAI models. Inaccurate or unreliable outputs, whether the anatomical visuals as exemplified in Fig~\ref{fig:teaser}A--D or blocks of code, can mislead and diffuse responsibility. Participants questioned who should be held accountable in instances where genAI is used and lines of accountability blur. As \dev{Kim} summarizes, \textit{``There should be someone who can explain the results. It is about trust, and [...] about competence.''} 
\dev{Ray} expands on this notion, stating that \textit{``it should not be something that's a black box and in the background. [You] need the corresponding user interfaces to establish what the visual mapping should be. [...] That's the danger, and it applies to all these stages. If you don't know what's going on and some magic happened, then potentially you're not aware of what's being shown, and you may miss something.''} \dev{George} adds that having \textit{``some tools to track the provenance of the data, like annotation tools or for data traceability''} would increase trust and explainability. 

Designers similarly feel this sense of responsibility toward accurate and explainable visual design. \comm{Isaac} notes, \textit{``We should stand behind the accuracy of it and we should be held responsible when we screw up.''} \comm{Ursula} similarly asserts, \textit{``You need to have the highest standards [...] for patient care. And patient education is 100\%. There is no room for error in this.''} Accountability is placed on the BioMedVis expert in medical legal visualization, as \comm{Arthur} explains: \textit{``As an expert in the field, you may be called to testify as to, how do you know this re-creation your studio has made is accurate? What is your background? How do you know this much about the subject? I don't think AI will ever have a place in that. Because how do you attest that some machine learning actually knew that particular person's hepatic artery had a branch in a place where a surgeon might have cut it accidentally?''}

While errors in genAI outputs might be obvious now, participants believe that these errors will become harder to catch as the technology improves and human users get more accustomed to trusting these systems. With a technology that cannot reach 100\% accuracy, participants are concerned that more critical errors will slip past human eyes. 

\textbf{BioMedVis experts advocate for diverse, fair representation; genAI models cannot.} 
Our participants express that part of the responsibility of the BioMedVis expert is to ensure, while working with respective stakeholders, that the visualizations they produce serve their intended purpose with sensitivity. Those involved in patient-centred projects raise repeated concerns about biases in training data that produce insensitive text and visual depictions of health or medical data, or limited opportunities for diverse representation. \dev{Martha} asserts, \textit{``In medicine, most things in the past were done with a man as a standard human. So it's also important to be mindful of diverse groups. It can be gender, but also ethnicity, and different regions. It's about trust, this is a big topic for me.''} \comm{Isaac} further stresses the need for built-in correctional safeguards against \textit{``unjust, historical mistreatment''} of minority groups, to build trust and achieve fair representation.

%
\subsection{The creative choices of the BioMedVis expert balance expression with scientific accuracy and purpose.}

\textit{``If it's not accurate, especially if it's done by someone [who's] a fantastic artist, but they don't have the science background to know that DNA is twisted the wrong way [...] it looks really sexy, but it's wrong---that's a big concern.''} 
\comm{Mary} captures the confluence of skills that a BioMedVis expert applies to the development of a biomedical visualization. Specific design choices (how), even if aesthetic, are modulated by the purpose (why/for whom) and the data being visualized (what) (see \cref{fig:triadframework}). \textit{``With the scientific side of it, I think there is a bit of a difference between creating something that is meant to be truly artistic and something that is meant to be based on science,''} asserts \comm{Margaret}.

\textbf{BioMedVis has a ``bias toward novelty'' which precludes genAI use beyond superficial ideation and implementation.}
All our participants agree that their work often involves developing unique, bespoke biomedical visualization solutions, especially to convey cutting-edge research. Additionally, the kinds of problems and data that BioMedVis experts work with are increasingly complex. \dev{Alan} uses the metaphor of trying to solve for \textit{``something between a nail and a screw, and something else,''} meaning that visualization solutions tend to be as unique as the problems they solve. GenAI models may be able to remix training data, but our participants believe true innovation and creativity still lies within the human creator, as \dev{George} expresses: \textit{``If you want something original [...] that's still on us, on the human side.''}

The aesthetic quality of AI-generated images is seen as simultaneously overwrought and generic, according to seven participants, and cannot make subtle, carefully reasoned design choices, such as where to add more or less detail for visual emphasis and clear messaging. Per \comm{Neil}: \textit{``By understanding the science, we know what the limits are that we can work within, we can maximize the aesthetic within those boundaries, still making accurate artwork but also making it beautiful and information-efficient, from a storytelling perspective.''}

Nevertheless, some participants see the utility of genAI tools for ideation, as previously covered in our workflow analysis. They are cognizant, however, of the risk of becoming overly reliant on this external ideation process without further independent exploration, and in particular for trainees or those newer to the field. \dev{Alan} and \comm{Jeff} emphasize that not only does one risk stifling creative thinking capabilities, it may limit their ability to develop novel solutions for cutting-edge research.


Ultimately, the perspectives shared by participants about genAI in the BioMedVis pipeline embody not just attitudes related to this technology but also \textbf{core values and concerns about their work that transcend genAI}. All participants share a commitment to scientific accuracy as one of the central concerns in BioMedVis, although developers express this commitment more in relation to the data while the designers express this as a commitment to the patient or other intended target audience. This latter perspective is reinforced under the theme relating to human-centred and empathetic design---here, we see the same values but articulated through different motivations. Values expressed by developers under this theme relate more to issues of trust and judgment around the data, while designers again primarily relate use and accountability to human as opposed to data concerns. Values related to creative expression, however, are similarly-rooted for both groups: designers and developers place high value on aesthetics, story, and novelty in their work, whether in developing novel shading algorithms, molecular dynamics animations, or didactic, painterly anatomy. 


\section{Discussion}
\label{sec:disc}

In this section, we synthesize findings from our respective pipeline and thematic analyses to explore the value landscape that we see underlying participants' overarching attitudes towards genAI. This value landscape affords deeper understanding as to why some participants find genAI useful in certain instances while others do not. 
The uses of genAI in the BioMedVis workflow, in some cases, affirm the speculations posed by Schetinger et al.~\cite{schetingerdoom2023}, while other instances are new---this is due largely to the new considerations introduced by biomedical data and our designer group of interviewees, whose standard output is more figurative than procedural. Our work continues the conversation within the community to understand possibilities and contexts of genAI use, alongside the tangible impacts it exerts on the everyday workflow for BioMedVis and visualization more broadly. 

The designers and developers we interviewed in our study are, in different ways, using or actively avoiding genAI tools at several points in the BioMedVis pipeline. We consider what the visualization community can learn from these points of disjointed use---where visualization research can support designers, and where visualization research may draw inspiration from designers in a genAI-supported BioMedVis workflow. 
We furthermore draw attention to the (often unseen or unacknowledged) labour and economic impact of genAI tools 
with a call-to-action to develop supporting structures for using genAI in ethical, expert-validated visual science communication. 
%

\subsection{Exploring participants' genAI value landscape}
\label{sec:adoptionspace}

\begin{figure}[ht]
  \centering
\includegraphics[width=0.95\linewidth]{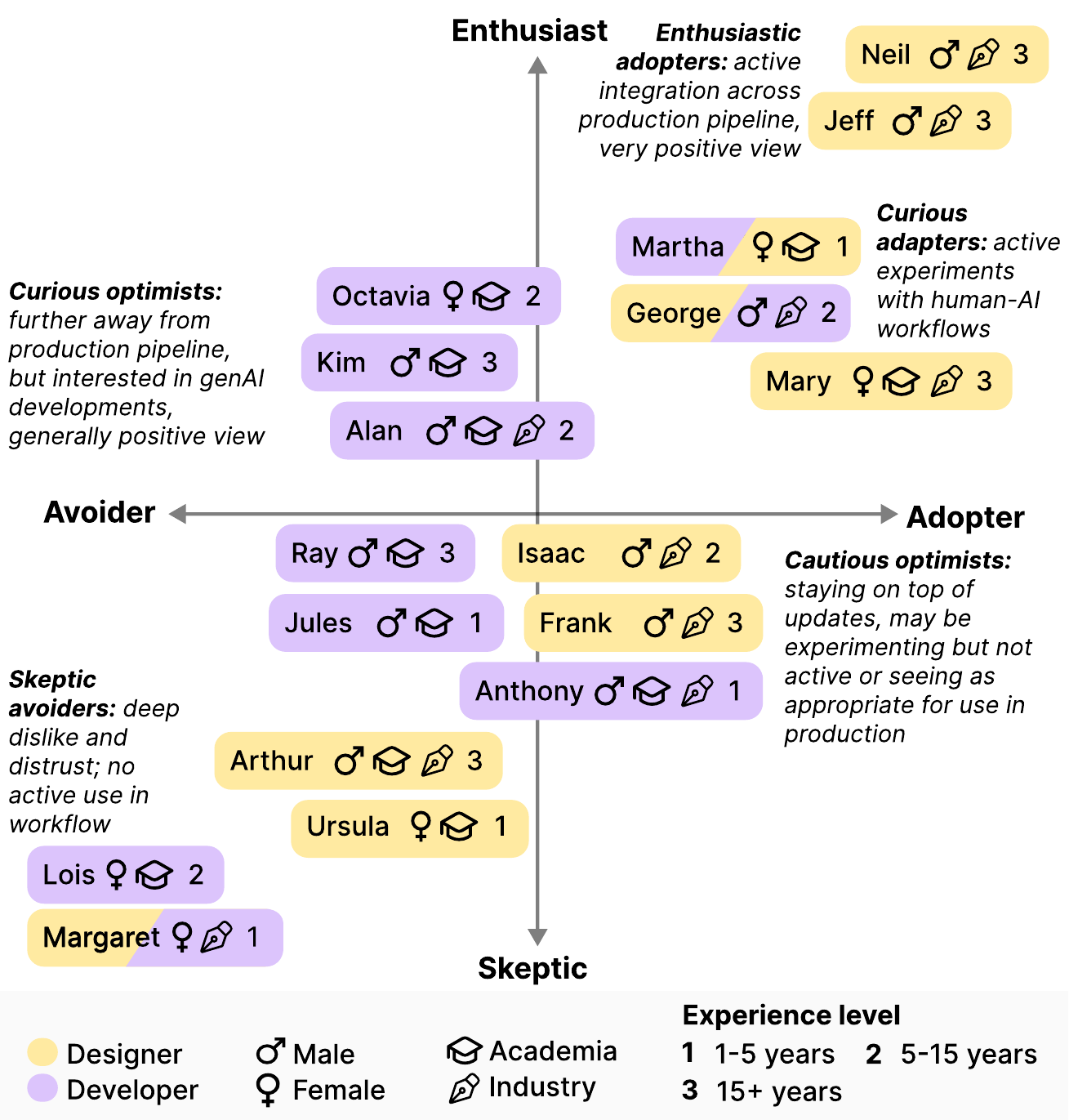}
  \caption{Participant profiles according to genAI attitudes (enthusiast $\updownarrow$ skeptic; active use (adopter $\longleftrightarrow$ avoider); positions are localized relative to their centroid. 
  Demographic information is also shown (\comm{designers} in yellow and \dev{developers} in purple, and split by: gender, professional setting, and experience level). Pseudonyms are given to protect identities. \revision{Three participants who have aspects of both roles (Martha, George, and Margaret) are referred by their self-identified primary role.}
  }
  \label{fig:profiles}
\end{figure}

Characterizing participants in a value landscape for genAI use, we find that their attitudes and active use varied greatly. Participants localize as \textit{enthusiasts} to \textit{skeptics} (how participants think or feel about genAI) and \textit{adopters} to \textit{avoiders} (how participants are actually using genAI tools) according to several factors: perceived genAI utility in their personal workflows, personal values, e.g., openness to using this technology, negative or positive attitudes towards genAI in general, and whether they willingly choose to engage with the technology or rather feel pressured to. Other aspects such as work context and experience play a role, but differently than we expected: the biggest skeptics range from early-career industry professional to late-stage academic, and the strongest adopters are in middle- to later-stage careers in both academia and industry. Here, we outline and reason about the varied attitudes we observed amongst our participants, arrayed into five personas: \textbf{Enthusiastic adopters}, \textbf{Curious adapters}, \textbf{Curious optimists}, \textbf{Cautious optimists}, and \textbf{Skeptical avoiders} \revision{(as shown in \cref{fig:profiles})}.

\textbf{Enthusiastic Adopters} 
(\comm{two senior-level})
\textbf{: } 
%
These participants draw from long experience in the field coupled with techno-positivist attitudes that we see resulting in their primary focus on the potential, rather than problems, of genAI. These attitudes towards technology help explain their finding a creative outlet for personal use of genAI, and their belief in this as a craft in its own right. Their techno-positivism, paired with their professional seniority that places them further from the production trenches, contextualizes their shared perspective as creative directors or conductors that guide the genAI tool to achieve their vision. Finally, we see their positioning within industry, as opposed to academia, as fueling a desire to stay competitive as technologies rapidly evolve. 

\textbf{Curious Adapters} 
(\dev{one junior-}, \comm{one mid-, one senior-level})
\textbf{: } 
Less enthusiastic but still inclined towards adoption are the BioMedVis experts who are slowly adapting, by actively exploring human-genAI hybrid workflows. Similar to the previous group, we see an inclination toward techno-positivism. Their curiosity and openness to experimentation is tempered by concerns around issues of provenance and equality.

\textbf{Curious Optimists} 
(\dev{two mid-, one senior-level})
\textbf{: }
This group comprises BioMedVis experts who express similar values of techno-positivism, curiosity, and openness to experimentation. However, they generally have a deeper understanding of computer science and the architecture of LLMs, and so are more reluctant to engage with tools with such known limitations. These participants also tend to be more distant from direct implementation tasks in their day-to-day work, and so may also be less tempted by genAI's promises of efficiency and productivity than others more directly involved in implementation. 

\textbf{Cautious Optimists} 
(\dev{three}, \comm{two}, mixed-level)
\textbf{: } 
Participants who cluster around the middle of the enthusiast/skeptic axis have a balanced view toward genAI. We see their moderate attitudes on genAI as ``just'' another tool, with its corollary risks and benefits, as a reflection of a similarly moderate view of technology's role in society. We place them nearer to the skeptics axis in consideration of their advocacy for specialized expertise in visually communicating science, reflecting a confidence in and valuing of their own expertise, while also staying current to changes in the field.

\textbf{Skeptical Avoiders} 
(\comm{one senior-, two junior-level}, \dev{one mid-level})
\textbf{: } 
For the skeptical avoiders, genAI is neither interesting nor useful. We do not see their frustration, distrust, and disdain toward the technology stemming from an anti-technology view, but rather see their conservative views towards genAI as a reflection of backgrounds that train the individual to be cautious and risk-averse. We see their disuse of the tools stemming in large part from a deep sense of pride in their learning and experience to become BioMedVis experts. This group also acknowledge they have close contacts in creative fields whose livelihood they have witnessed already threatened by genAI. Here, we also see indications that experience may actually lead to negative feelings towards genAI--this group is mostly early career-stage. 


All participants described a \textit{need for control} throughout the BioMedVis pipeline, but this control looks different depending on their attitude toward genAI. The \textbf{skeptical avoiders} reject genAI entirely, as they deem the risks to be too great. We also see a strong sense of ownership over their knowledge and cultivated skills that they do not feel the need to `augment.' The \textbf{cautious optimists} still very much see their own direct, hands-on work as being central to their process, and feel encumbered by having to correct genAI outputs. They also question what would happen to their sense of creative fulfillment and joy were they to offload `lower-level' tasks, like basic coding or rendering, to genAI. The \textbf{curious optimists} and \textbf{curious adapters} likewise value their craft, but are ready to adapt their workflows to use genAI for lower-level tasks or, interestingly, to generate content that is outside their core competency (e.g., sound, character design). The \textbf{enthusiastic adopters} describe their sense of control akin to that of a conductor, and so see genAI as a value add to their work.

\subsection{Comparing the speculated vs. current landscape of genAI in the visualization pipeline}

\begin{figure}
  \centering
  \includegraphics[width=\linewidth]{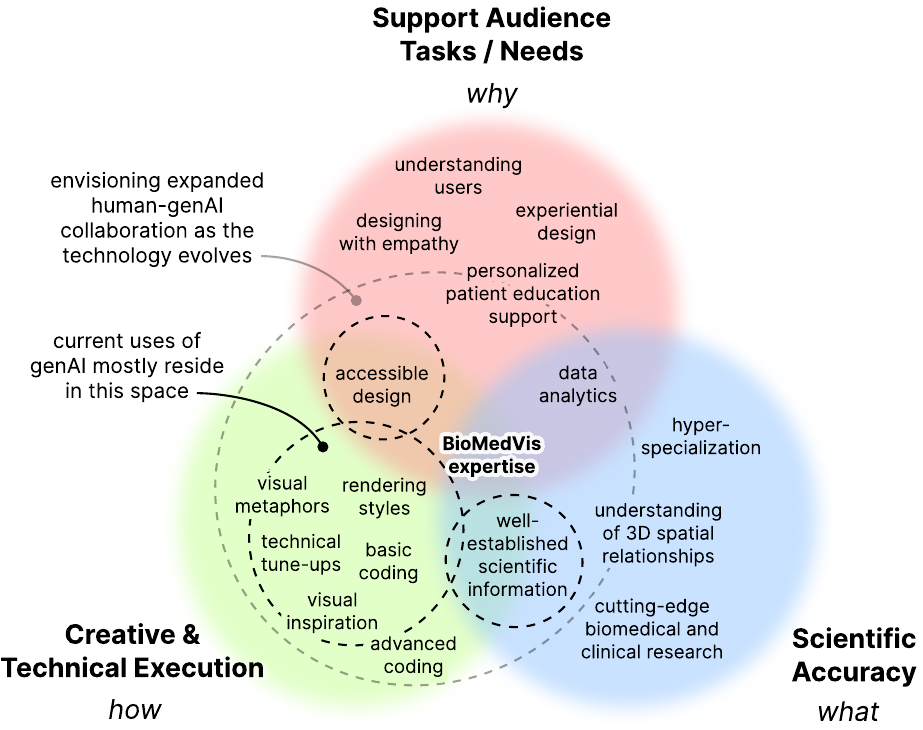}

  \caption{The overarching themes of scientific accuracy, supporting audience tasks/needs, and creative/artistic license, with specific workflow elements, capture BioMedVis expert perspectives on genAI use in their daily work. These themes represent a tri-lens perspective on the boundaries of ethical AI use in current practice. \revision{Dotted lines indicate envisaged bounds of genAI use (\textit{black = current; grey = possible)}.} 
  }
  \label{fig:triadframework}
\end{figure}

GenAI tools are developing rapidly to alter the visualization pipeline. To illustrate these changes and their effects on individuals in this field, we contrast study participants' experiences with Schetinger et al.~\cite{schetingerdoom2023}'s speculations 
in their 2023 study.
\cref{fig:triadframework} captures our participants' perspectives on \textit{current, possible,} and \textit{unlikely} genAI use in three areas: scientific accuracy for \textit{what} is being visualized (blue); 
the task/need, or \textit{why}, for developing a new method for, or producing, a visualization (red); 
and \textit{how} creative choices drive the mechanics of visualization production (green). In each area, we discuss opportunities for collaboration with designers situated outside the core visualization field. 


\textbf{Contrasting speculated vs. current support for scientifically accurate visuals. }
Schetinger et al.~\cite{schetingerdoom2023} speculated that genAI could support several tasks related to capturing and transforming data \revision{while noting concerns about trustworthiness and bias in the outputs.
Our participants echoed these concerns, indicating infrequent use of genAI for such tasks beyond basic data clean-up, and so the tension between convenience and trust is ever-present. Additionally, \textit{the need for precise rendering of 3D relationships in BioMedVis} limits the accuracy of genAI outputs.} On the other hand, Schetinger et al.~\cite{schetingerdoom2023}'s posit that genAI may be used to synthesize new data was confirmed by one of our participants to generate variant anatomy. \revision{As models improve, we anticipate that with thoughtful parameter-space guidance, designers and developers may create imagery of structural variants of anatomy to facilitate targeted and personalized clinical training. }

\textbf{Contrasting speculated vs. current support for creative \& technical execution.} 
\revision{In our study, we discussed beyond Schetinger et al.~\cite{schetingerdoom2023} to the careful \textit{balance} between creative expression and scientific accuracy, and the need for the BioMedVis expert to retain \textit{control} and \textit{oversight} as scientific translator and visual storyteller. Interfaces that allow BioMedVis experts more fine-grained tuning of regions of genAI output would be helpful in maintaining this balance.

Schetinger et al.~\cite{schetingerdoom2023} further predicted tasks related to visual design, including moodboarding, basic coding, rapid prototyping, and stylization or ``beautification,'' which were echoed in our study, although with a caveat: \textit{in an industry where unique style is key to winning client contracts, the bias toward novelty in BioMedVis generally limits the perceived utility of genAI, especially for many designers.}  

Copyright concerns regarding AI-generated images were mixed in Schetinger et al.~\cite{schetingerdoom2023}'s study. By contrast, BioMedVis designers \textit{do} see their output as artistic artifacts, and so have a stronger sense of how genAI and copyright risk impact their work. As such, they offered \textit{a clearer delineation for appropriate use of AI-generated elements in the pipeline, i.e., in earlier stages of development, and in later stages only if minor or modified.} Tooling that assists designers in labelling when and how genAI is used in the pipeline would help address this issue.} 


\textbf{Contrasting speculated vs. current support for design with nuance. }
\revision{Between our two studies, design for accessibility is seen as a meaningful application of genAI and is an active area of research (e.g.,~\cite{yanez2024adaptive}). In BioMedVis, this benefits patient education and public health outreach.
\textit{The challenge 
remains in assembling multiple elements (text, visuals, interactivity) into comprehensive pieces that communicate information effectively---a task for which genAI, according to our participants' experience, is not ready.} 

While Schetinger et al.~\cite{schetingerdoom2023} focused on the nuts and bolts of the data visualiation pipeline, a large part of the interviews in our study highlighted another aspect central this work: human-human collaboration.
Our participants affirm that the social dimensions are equally essential as the scientific~\cite{gawande2008}. They emphasize that health and medical data must be conveyed with care and sensitivity---which is best done through directly interfacing with multiple stakeholders and partners, be they clinicians, researchers, patients, students, or general audiences. \textit{Empathy is an innately human characteristic, and to design with empathy requires understanding and balancing multiple stakeholders' views, which, our participants assert, the human creator best poised to do.}}
\textit{Situating genAI use in a specialized area of visualization that balances the fine line between science and figurative art, we see the essential role of human agents to tell scientific stories with intention}, 
echoing calls for a human-in-the-loop in the visual analytics pipeline~\cite{Keim2008}.


\subsection{A call to action for science communication}

\revision{Thoughtful, nuanced design serves as an essential bridge to how our target audiences receive scientific content. }
In contemplating how we collectively reason about genAI, we frame human partnerships with genAI tools as an exercise in \textit{crafting}, rather than finding or revealing, data~\cite{walford_data_2020}. This process of data crafting includes numerous touch points where we make creative \textit{choices} and use contextual knowledge to produce a result that reflects our understanding of reality. Using genAI to facilitate the crafting of data and visuals eliminates many of these reflective opportunities for \textit{human choice} in favour of \textit{machine decision}, resulting in a synthesized product made of code and pixels that cannot be fully rationalized. For BioMedVis, this is problematic since spatial relationships and entangled variables require fine-grained rationalization to be considered trustworthy, issues which can extend into the broader space of science communication when genAI tools are wielded to \textit{craft decisions} rather than through \textit{careful choices}. 

\revision{The current landscape of genAI attitudes and use in the BioMedVis pipeline relative to prior speculations affords several opportunities for reflection on its role in BioMedVis and visual science communication more broadly. We focus key takeaways from our study in two areas: (1) the need for provenance and validation of genAI scientific and stylistic output to facilitate adoption, particularly for designers, and (2) the necessity of ethically sourcing training data that are truly representative of a target group for equitable science communication.}

\revision{Provenance and validation tools for genAI outputs, including scientific and stylistic elements, are necessary to improve BioMedVis experts' adoption of genAI. Close collaboration and prototyping with designers will be essential to understand the numerous touch points where these validation tools will be most useful in crafting data, and managing sources~\cite{jantzen2015transparency} in BioMedVis.
There are other opportunities for visualization research to support genAI-assisted workflows, e.g., creating interfaces to semi-automate visual metaphor design and visual inspiration for moderately-complex anatomical structures. Data interpolation for medical animations (e.g., similar to~\cite{cascadeur_casc_nodate}), or interpolation of 3D spatial data for more common structures, are also possible applications. Visualization research has drawn inspiration from illustrative visualization in years past, and we see this continuing through, e.g., style transfers for rendering anatomy, or in narrative visualization~\cite{segel2010narrative, meuschke_narrative_2022} to translate visual styles according to different audiences for data-driven medical stories. Importantly, these must be steered by a human overseeing an ethically-trained model.
}

\revision{Ethical sourcing of \textit{truly} representative training data is necessary for equitable visual science communication.} This addresses current underlying societal biases in genAI models that further propagate harmful stereotypes~\cite{omiye_large_2023, correll2024bodydata}, and reflect an underlying human problem~\cite{scheuerman_products_2024}. This is particularly problematic in medicine, a domain with a record of poor treatment of diverse groups. In recent years, the BioMedVis field has been reckoning with its own history of under-representation in medical illustration, with initiatives to remediate these historic biases and improve visibility for diverse groups (e.g.~\cite{fung_skin_2024}; \textit{Illustrate Change}~\cite{association_of_medical_illustrators_ami_2023,johnson__johnson_illustrate_2023}), thereby promoting better clinical training and patient education. Similar efforts to improve diversity are being explored in broader visualization research~\cite{dhawka_we_2023, dhawka_better_2024, cabric_eleven_2023}. Such efforts are unlikely to be driven or easily facilitated by genAI since, currently, the overwhelming majority of training data available for BioMedVis remain white, fit, and male~\cite{correll2024bodydata}. Issues of representation naturally surface broader questions of trust. We ask, then, how genAI technology interweaves with contemporary sociocultural concerns that impact the meaning and value of BioMedVis and other creative work~\cite{correll2024bodydata}. 
How we drive dialogue about these very issues could lead to policy changes for more ethically-sourced training data to promote equitable communication of science. In many ways, the BioMedVis community serves as a conduit to facilitate, or destabilize, public trust in science through our work.

\revision{As genAI technologies integrate more seamlessly into creative and professional workflows, we find ourselves operating within an expanding grey area concerning what constitutes ethical use. How should we properly credit genAI when it contributes to our work? Conversely, is there a need to explicitly state when genAI has not been used? These questions are not just theoretical; they have real implications for how work is evaluated, both in terms of value and authenticity. For both developers and designers, this shift invites a reconsideration of how their work is perceived and valued. If part of the labour is shared with a genAI tool, how might our effort, originality, and worth of that output be assessed? Critical reflection of the ethical and legal boundaries within which we operate is essential as we move forward.}

Major public health events like the COVID-19 pandemic demonstrate the importance, and consequences, of \textit{timely and reliable} communication of medical and health data. 
Challenging this is the free availability of genAI tools, which has enabled the creation of eye-catching visual content at mass scale and allows anyone to create and share potentially misleading content, intentionally or otherwise. 
We must consider the consequences of genAI use for public education, engagement, and trust in science, particularly via online channels~\cite{klein2024make}. We must advance research that investigates the trustworthiness of visualizations produced through hybrid human-AI workflows, as these consequences reach beyond BioMedVis. News outlets regularly include data-driven graphics with spatial information for weather or geography, and often encode multiple, intertwined variables in complex infographics. Careful action at a policy-level is needed to validate possibly misleading or harmful visuals. We call on the broader visualization community to contribute to this effort by formalizing and strengthening data provenance practices (e.g., similar to the use of `Data nutrition labels'~\cite{chmielinski_clear_2024}), alongside annotation and validation methods that can both assist communication designers in developing visualizations, and enable audiences to understand the data sources and process that informed the development of a visualization~\cite{jantzen2024design,liu2022annotating}, including the use of genAI.

The networks between science and society are complex~\cite{kupper_rethinking_2021}, with the public having access to multitudes of conflicting sources of scientific information. This problem is not new; it has taken on various guises over the years through the advent of new technologies and shifting attitudes across all branches of science. Now, with the ability for virtually anyone to create and shape convincing stories that influence scientific discourse and pursuit, we reflect on our imperative to visually communicate science with integrity, to \textit{first do no harm}, as articulated by Goodsell \& Johnson~\cite{goodsell_filling_2007}:
\textit{``As communicators who make science accessible to colleagues, students, and the public, \textbf{we must define an acceptable licensing threshold that will allow us to create evocative pictures, but will still leash us enough to avoid polluting the literature (both scientific and popular) with deceptive imagery.} This is particularly important for scientists, since pictorial errors in primary scientific publications, which are often persistent sources of information, may be propagated for decades in educational and outreach publications.''} 
%


\section{Study Limitations and Future Work}
In this qualitative interview study, we explored the perspectives of 17 BioMedVis experts toward genAI in the field. We acknowledge our limited focus within the Euro-American context, and propose further work to understand genAI's impact on the field in other geographic and cultural contexts. 
Additionally, genAI is already shaping BioMedVis education and training, and although this topic was of great interest to our participants and us, it fell outside the scope of this present work. 

We focused our questions on participants' use of text-to-text and text-to-image prompt-based genAI tools publicly available and current through early 2025, including ChatGPT, GitHub Copilot, MidJourney, and DALL-E. We remain curious about future shifts in attitudes and adoption. For instance, ours and similar studies have uncovered little about environmental concerns~\cite{crawford2021atlas} with genAI use, or the hidden human labour that goes into training large genAI models~\cite{crawford2021atlas}. As awareness about these costs increases, future work remains to understand its impact in genAI adoption across BioMedVis and other fields.

We did not interview the receivers of AI-generated biomedical imagery. Logical follow-on work to creators' perspectives is to assess audience perceptions, to learn how their understanding and trust in biomedical topics and science more broadly are affected by the profusion of beautiful, but inaccurate, scientific imagery.

\section{Conclusion}
BioMedVis lies at the tension of science and figurative art, drawing from diverse disciplines to create visuals for health and medicine. As the latest disruptive technological innovation in the history of computing, genAI is poised to both support and hamstring this field. GenAI use is mediated by the BioMedVis field's core commitment to scientific accuracy and visualizing the human body and its mechanisms with empathy. \revision{Concerns such as the way one and their work is perceived, the nature of work itself in this hyper-specialized field, and bottom-line financials, drive individual attitudes and (dis)use of genAI tools. Primarily used as an auxiliary workflow tool now, we find that, e.g., career stage, personal philosophies towards technology, and a deep sense of responsibility shape whether genAI is seen to \textit{craft decisions} or facilitate \textit{careful} human \textit{choices} in service of scientific storytelling.} In comparing current genAI use to earlier speculations, we continue the conversation around genAI as an assistive or disruptive agent for visualization. We further assert the need for continuous human intervention in a pipeline with potential for high-stakes consequences, should a ``sexy'' visual misrepresent essential health information. Opportunities for visualization research to support these aims and enhance genAI-assisted workflows can enhance broader adoption by the design community in particular. However, equally important to BioMedVis work are the relational, human aspects, which should not be lost in our efforts toward accurate, trustworthy, and engaging scientific visuals.

\section*{Supplemental Materials}
\label{sec:supplemental_materials}
Supplemental materials including study artifacts are available on OSF at \href{https://osf.io/mbw86/?view_only=7710ece36e1a4893a97f1e0377cc6360}{\texttt{https://osf.io/genaixbiomedvis/}}, released under a CC BY 4.0 license. 
They include: (1) Manuscript figures; (2) Table 1 of study comparisons; (3) Workflow and theme analysis (anonymized): Screenshots from our \href{https://miro.com/app/board/uXjVLdKIqFQ=/}{Miro board} with descriptions that illustrate and detail our analysis process; (4) Interview guide and note-taking template; (5) Full participant quotations included in the manuscript; (6) Additional supplementary figures.

\section*{Figure Credits}
\label{sec:figure_credits}
\textit{\cref{fig:teaser}A:} DALL-E 3 prompt: Medical illustration of the inferior view of the skull; \textit{\cref{fig:teaser}B:} DALL-E 3 prompt: Medical illustration of the knee joint from the 3/4 view showing range of motion in medial-lateral, anterior-posterior and superior-inferior axis; \textit{\cref{fig:teaser}C:} DALL-E 3 prompt: Illustration of a cell, with transparency/ghosting to show internal structures; \textit{\cref{fig:teaser}D:} DALL-E 3 prompt: fMRI showing neuronal activity in the left motor cortex of the brain; \textit{\cref{fig:teaser}E:} Reused with permission by Roxanne Ziman; \textit{\cref{fig:teaser}F:} Reused with permission by Shehryar Saharan. \textit{\cref{fig:teaser}G:} Reused with permission by Digizyme and Shehryar Saharan. \textit{\cref{fig:teaser}H:} Reused with permission Solteszova et al.~\cite{Solteszova-2019-MLT}. We, as authors, state that all original figures and works produced by us are and remain under our own personal copyright, with permission to be used here. We also make them available under the Creative Commons Attribution 4.0 International 
\href{https://creativecommons.org/licenses/by/4.0/}{(CC BY 4.0)} license and share them at \href{https://osf.io/mbw86/?view_only=7710ece36e1a4893a97f1e0377cc6360}{\texttt{https://osf.io/genaixbiomedvis/}}. 

\acknowledgments{
We are grateful to our participants for their time and invaluable insights that made this project possible. We also thank Michael Correll, Tobias Isenberg, Kevin Doherty, Jodie Jenkinson, Jan Byška, Jill Rettberg, and our reviewers for thoughtful discussions and feedback on drafts of this manuscript. This work was supported by the Univ. of Bergen and Trond Mohn Foundation (\#813558).%
}

\bibliographystyle{abbrv-doi-hyperref}

\bibliography{__references_CLEANED}

\appendix

\end{document}